\documentclass[aps,prd,twocolumn,superscriptaddress]{revtex4}
\usepackage{epsfig,epsf}
\usepackage{amsmath}
\usepackage{amsthm}
\usepackage{amsfonts}
\usepackage{amssymb}
\usepackage{dsfont}
\usepackage{multirow}
\usepackage{appendix}
\usepackage{slashed}
\usepackage[active]{srcltx}
\usepackage{psfrag}

\begin{document}

\title{Resonance $Y(4660)$ as a vector tetraquark and its strong decay
channels}
\date{\today}
\author{H.~Sundu}
\affiliation{Department of Physics, Kocaeli University, 41380 Izmit, Turkey}
\author{S.~S.~Agaev}
\affiliation{Institute for Physical Problems, Baku State University, Az--1148 Baku,
Azerbaijan}
\author{K.~Azizi}
\affiliation{Department of Physics, Do\v{g}u\c{s} University, Acibadem-Kadik\"{o}y, 34722
Istanbul, Turkey}
\affiliation{School of Physics, Institute for Research in Fundamental Sciences (IPM),
P.~O.~Box 19395-5531, Tehran, Iran}

\begin{abstract}
The spectroscopic parameters and partial widths of the strong decay channels
of the vector meson $Y(4660)$ are calculated by treating it as a bound state
of a diquark and antidiquark. The mass and coupling of the $J^{PC}=1^{--}$
tetraquark $Y(4660) $ are evaluated in the context of the two-point sum rule
method by taking into account the quark, gluon and mixed condensates up to
dimension 10. The widths of the $Y(4660)$ resonance's strong $S$-wave
decays to $J/\psi f_{0}(980)$ and $\psi(2S) f_{0}(980)$ as well as to $%
J/\psi f_{0}(500)$ and $\psi(2S) f_{0}(500)$ final states are computed. To
this end, strong couplings in the relevant vertices are extracted from the
QCD sum rule on the light cone supplemented by the technical methods of the
soft approximation. The obtained result for the mass of the resonance $%
m_Y=4677^{+71}_{-63}\ \mathrm{MeV}$ and the prediction for its total width $%
\Gamma_{Y}= (64.8 \pm 10.8)\ \mathrm{MeV}$ is in nice agreement with the
experimental information.
\end{abstract}

\maketitle


\section{Introduction}

\label{sec:Inrtoduction}
The last 15 years were very fruitful for hadron physics due to valuable
information on properties of the hadrons collected by numerous experimental
collaborations and owing to new theoretical ideas and predictions that
extended boundaries of our knowledge about the quark-gluon structure of
elementary particles. An observation of the resonances that may be
interpreted as four- and five-quark states is one of most interesting
discoveries to be mentioned among these achievements. Strictly speaking, the
existence of the multiquark states does not contradict the fundamental
principles of QCD and was foreseen in the first years of QCD \cite%
{Jaffe:1976ig}, but only results of the Belle Collaboration about the narrow
resonance $X(3872)$ placed the physics of multiquark hadrons on a firm basis
of experimental data \cite{Choi:2003ue}. Now experimentally detected and
theoretically investigated four-quark resonances form a family of  particles
known as $XYZ$ states \cite{Chen:2016qju,Esposito:2016noz}.

The resonance $Y(4660)$ which is the subject of the present study was
observed for the first time by the Belle Collaboration in the process $%
e^{+}e^{-}\rightarrow \gamma _{\mathrm{ISR}}\psi (2S)$ $\pi ^{+}\pi ^{-}$
via initial-state radiation (ISR) as one of two resonant structures in the $%
\psi (2S)\pi ^{+}\pi ^{-}$ invariant mass distribution \cite%
{Wang:2007ea,Wang:2014hta}. The second state discovered in this experiment
received the label $Y(4360)$. The analysis carried out in Refs.\ \cite{Wang:2007ea,Wang:2014hta} showed that
these structures cannot be interpreted as known charmonium states. The
measured the mass and total width of the resonance $Y(4660)$ are \cite%
{Wang:2014hta}
\begin{eqnarray}
m_{Y} &=&4652\pm 10\pm 8\ \mathrm{MeV},  \notag \\
\Gamma _{Y} &=&68\pm 11\pm 1\ \mathrm{MeV}.  \label{eq:SpPar}
\end{eqnarray}

The state $Y(4630)$, which is usually identified with the $Y(4660)$, was
detected in the process $e^{+}e^{-}\rightarrow \Lambda _{c}^{+}\Lambda
_{c}^{-}$ as a peak in the $\Lambda _{c}^{+}\Lambda _{c}^{-}$ invariant mass
distribution \cite{Pakhlova:2008vn}. Making an assumption on a resonance
nature of this peak it mass and width were found equal to $%
m_{Y}=4634_{-7}^{+8}(\mathrm{stat.})_{-8}^{+5}(\mathrm{sys.})\ \mathrm{MeV}$
and $\Gamma _{Y}=92_{-24}^{+40}(\mathrm{stat.})_{-21}^{+10}(\mathrm{sys.})\
\mathrm{MeV}$, respectively. Independent confirmation of the $Y(4660)$ state
came from the BaBar Collaboration \cite{Lees:2012pv}, which studied the same
process $e^{+}e^{-}\rightarrow \gamma _{\mathrm{ISR}}\psi (2S)\pi ^{+}\pi
^{-}$ and fixed two resonant structures  in the $\pi ^{+}\pi ^{-}\psi (2S)$
invariant mass distribution. Their mass and width confirm that these
structures can be identified with resonances $Y(4660)$ and $Y(4360)$.
Besides two resonances under discussion there are also states $Y(4260)$ and $%
Y(4390)$ which together constitute the family of at least four $Y$
hidden-charmed particles with $J^{PC}=1^{--}$.

The numerous theoretical articles claiming to interpret the $Y(4660)$ and $%
Y(4360)$ embrace variety of models and schemes available in high-energy
physics. Thus, attempts were made to consider the new resonance $Y(4660)$ as
an excited state of conventional charmonium: in Refs.\ \cite{Ding:2007rg}
and \cite{Li:2009zu} it was interpreted as the excited $5{}^{3}S_{1}$ and $%
6{}^{3}S_{1}$ charmonia, respectively. To explain the experimental
information on the resonance $Y(4660)$ it was examined as a compound of the
scalar $f_{0}(980)$ and vector $\psi (2S)$ mesons \cite%
{Guo:2008zg,Wang:2009hi,Albuquerque:2011ix}, or as a baryonium state \cite%
{Qiao:2007ce,Cotugno:2009ys}. The hadrocharmonium model for these resonances
was suggested in Ref.\ \cite{Dubynskiy:2008mq}.

The most popular models for the states $Y(4360)$ and $Y(4660)$, however are
the diquark-antidiquark models which suggest that these resonances are
tightly bound states of a diquark and an antidiquark with required quantum
numbers. Within this picture the resonance $Y(4360)$ was analyzed in Ref.\
\cite{Ebert:2008kb} as an excited $1P$ tetraquark built of an axial-vector
diquark and antidiquark, whereas $Y(4660)$ [and also $Y(4630)$] was found to
be the $2P$ state of scalar diquark-antidiquark. Calculations there were
carried out in the context of the relativistic diquark picture. The
resonance $Y(4360)$ was interpreted as a radial excitation of the tetraquark
$Y(4008)$ in Ref.\ \cite{Maiani:2014}. A similar idea but in the framework of
the QCD sum rule method was realized in Ref.\ \cite{Zhang:2010mw}: the $%
Y(4660)$ was considered as the $P$-wave $[cs][\overline{c}\overline{s}]$
state and modeled by $C\gamma _{5}\otimes D_{\mu }\gamma _{5}C$ type
interpolating current, where $C$ is the charge conjugation matrix. The
tetraquark $[cs][\overline{c}\overline{s}]$ with interpolating current $%
C\gamma _{5}\otimes \gamma _{5}\gamma _{\mu }C$ was used in Ref.\ \cite%
{Albuquerque:2008up} to treat $Y(4660)$, and the mass of this state was
evaluated by employing the QCD sum rule approach in  nice agreement with
experimental data. There are many other interesting models of the vector
resonances details of which can be found in the reviews (see Refs.\ \cite%
{Chen:2016qju,Esposito:2016noz}).

In general, the vector tetraquarks with different $P$ and $C$ parities can
be built using the five independent diquark fields with spin $0$ and $1$ and
different $P$-parities \cite{Chen:2010ze}. This implies the existence of
numerous diquark-antidiquark structures, and, as a result, different
interpolating currents with the same quantum numbers $J^{PC}=1^{--}$. Within
the framework of the two-point sum rule method these currents, excluding
ones with derivatives, were used in Ref.\ \cite{Chen:2010ze} for calculating
masses of the vector tetraquarks with $J^{PC}=1^{-+},1^{--},1^{++},\ 1^{+-}$
and quark contents $[cs][\overline{c}\overline{s}]$ and $[cq][\overline{c}%
\overline{q}]$. For the mass of the $1^{--}$ $[cq][\overline{c}\overline{q}]$
state, all of the explored currents led to the result $m\sim 4.6-4.7\
\mathrm{GeV}$, which implies a possible tetraquark interpretation of $Y(4660)
$. But this fact does not exclude interpretation of $Y(4660)$ as the state $%
1^{--}$ $[cs][\overline{c}\overline{s}]$, because the $C\gamma ^{\nu
}\otimes \sigma _{\mu \nu }C-C\sigma _{\mu \nu }\otimes \gamma ^{\nu }C$
type current gives for the mass of such a state $m=4.64\pm 0.09\ \mathrm{GeV}
$ comparable with the mass of the $Y(4660)$ resonance. The sum rule approach
was also employed in Refs.\ \cite{Wang:2013exa,Wang:2016mmg,Wang:2018rfw} to
investigate the resonance $Y(4660)$ by considering it  a tetraquark with $%
[cq][\overline{c}\overline{q}]$ or $[cs][\overline{c}\overline{s}]$ quark
content and using the interpolating currents of $C\gamma _{\mu }\otimes
\gamma _{\nu }C-C\gamma _{v}\otimes \gamma _{\mu }C$ and $C\otimes \gamma
_{\mu }C$ types.

In the present work we treat the $Y(4660)$ resonance as the vector
tetraquark with $[cs][\overline{c}\overline{s}]$ content and compute its
total width. To this end, we first recalculate the mass and coupling of $%
Y(4660)$, which enter as the important input parameters into its partial
decay widths. We utilize the two-point QCD sum rule approach, which is one
of the powerful nonperturbative methods for investigating the features of
the hadrons \cite{Shifman:1978bx,Shifman:1978by}. It is suitable for
studying not only conventional hadrons, but also multiquark systems. In our
computations we take into account vacuum condensates up to dimension 10,
which lead to reliable predictions for quantities of interest.

The next problem addressed in the present article is investigation of the $%
Y(4660)$ state's strong decays. Some of the possible decay channels of the
vector tetraquarks were written down in Ref.\ \cite{Chen:2010ze}. Our aim is
to evaluate the width of the main $S$-wave decays $Y\rightarrow J/\psi
f_{0}(980)$, $Y\rightarrow \psi (2S)f_{0}(980)$, $Y\rightarrow J/\psi
f_{0}(500)$ and $\ Y\rightarrow \psi (2S)f_{0}(500)$ of the resonance $%
Y(4660)$ and estimate its full width that can be confronted with existing
data. To this end, we employ the QCD sum rule on the light cone (LCSR) in
conjunction with a technique of the soft approximation \cite%
{Balitsky:1989ry,Belyaev:1994zk}. For investigation of the tetraquarks this
approach was adapted in Ref.\ \cite{Agaev:2016dev} and used successfully to
investigate their numerous strong decays.

This article is structured in the following manner: In Sec. \ref{sec:Mass}
we calculate the mass $m_{Y}$ and coupling $f_{Y}$ of the vector $Y(4660)$
resonance using the two-point sum rule method and include in the analysis
the quark, gluon and mixed condensates up to dimension 10. The obtained
results for these parameters are applied in Sec.\ \ref{sec:Decays} to
evaluate strong couplings and widths of the $Y(4660)$ state's partial $S$%
-wave decays. In Sec. \ref{sec:Conc} we present our conclusions. The
Appendix contains technical details of calculations.


\section{Mass and coupling of the vector tetraquark $Y(4660)$}

\label{sec:Mass}
In this section we revisit the sum rule calculation of the mass and coupling
of the resonance $Y(4660)$ to extract their values. In the context of the
QCD sum rule method this problem was originally addressed in Refs.\ \cite%
{Albuquerque:2008up,Chen:2010ze}, in which $Y(4660)$ was considered as the
state with $[cq][\overline{c}\overline{q}]$ or $[cs][\overline{c}\overline{s}%
]$ content. In these papers the relevant interpolating current was
constructed using different assumptions on quantum numbers of the
constituent diquark and antidiquark.

Here, we treat $Y(4660)$ as the $[cs][\overline{c}\overline{s}]$ tetraquark
composed of the scalar diquark and vector antidiquark with the $C\gamma
_{5}\otimes \gamma _{5}\gamma _{\mu }C$ type interpolating current. The same
assumption about the quark content and structure of the $Y(4660)$ resonance
was made in Refs.\ \cite{Albuquerque:2008up,Chen:2010ze}, in which its mass
was found by employing various interpolating currents and quark, gluon and
mixed vacuum condensates up to dimension 8. In our calculations we take into
account condensates up to dimension 10 and include in the analysis the gluon
condensate $\langle g_{s}^{3}G^{3}\rangle $ neglected in these papers and
improve accuracy of the obtained results. We do not restrict ourselves by
calculation of the mass of the resonance $Y(4660)$, as  was done in the
aforementioned works, and also extract the current coupling of the
tetraquark $Y(4660)$ which is necessary for investigating  its decay
channels.

After these preliminary comments, let us turn to our problem and start from
the analysis of the correlation function
\begin{equation}
\Pi _{\mu \nu }(p)=i\int d^{4}xe^{ipx}\langle 0|\mathcal{T}\{J_{\mu
}(x)J_{\nu }^{\dagger }(0)\}|0\rangle .  \label{eq:CF1}
\end{equation}%
Here $J_{\mu }(x)$ is the interpolating current of the resonance $Y(4660)$
chosen in the form
\begin{eqnarray}
&&J_{\mu }(x)=\epsilon \widetilde{\epsilon }\left[ s_{b}^{T}(x)C\gamma
_{5}c_{c}(x)\overline{s}_{d}(x)\gamma _{5}\gamma _{\mu }C\overline{c}%
_{e}^{T}(x)\right.   \notag \\
&&\left. +s_{b}^{T}(x)C\gamma _{\mu }\gamma _{5}c_{c}(x)\overline{s}%
_{d}(x)\gamma _{5}C\overline{c}_{e}^{T}(x)\right] ,  \label{eq:Curr1}
\end{eqnarray}%
where $\epsilon \widetilde{\epsilon }=\epsilon _{abc}\epsilon _{ade}$ and $%
a,\ b,\ c,\ d$ and $e$ are color indices.

In general, $\Pi _{\mu \nu }(p)$ has the  Lorentz decomposition
\begin{equation}
\Pi _{\mu \nu }(p)=\left( -g_{\mu \nu }+\frac{p_{\mu }p_{\nu }}{p^{2}}%
\right) \Pi _{\mathrm{V}}(p^{2})-\frac{p_{\mu }p_{\nu }}{p^{2}}\Pi _{\mathrm{%
S}}(p^{2}),  \label{eq:LorDec}
\end{equation}%
where the invariant functions $\Pi _{\mathrm{V}}(p^{2})$ and $\Pi _{\mathrm{S%
}}(p^{2})$ are contributions of the vector and scalar states, respectively.
Because we are interested only in the analysis of $\Pi _{\mathrm{V}}(p^{2})$%
, it is convenient to choose such a structure in Eq.\ (\ref{eq:LorDec}),
which accumulates effects due to only the vector particles. It is seen, that
such a Lorentz structure is $g_{\mu \nu }$; in fact, the terms proportional
to $p_{\mu }p_{\nu }$ are formed owing to both the vector and scalar
particles.

Deriving the sum rules for the mass $m_{Y}$ and coupling $f_{Y}$ proceeds
through two main stages. In the first step, we express the correlation
function in terms of the physical parameters of the tetraquark $Y(4660)$
which give rise to the function $\Pi _{\mu \nu }^{\mathrm{Phys}}(p)$. In the
next phase, we employ the explicit expression of the interpolating current $%
J_{\mu }(x)$, and calculate $\Pi _{\mu \nu }(p)$ contracting relevant quark
fields and replacing the obtained propagators with their nonperturbative
expressions. As a result of these manipulations, we get $\Pi _{\mu \nu }^{%
\mathrm{OPE}}(p)$, which depends on the various quark, gluon and mixed
vacuum condensates. By invoking assumptions about the quark-hadron duality
we can equate the functions $\Pi _{\mu \nu }^{\mathrm{Phys}}(p)$ and $\Pi
_{\mu \nu }^{\mathrm{OPE}}(p)$ to each other, fix invariant amplitudes
corresponding to the chosen Lorentz structure, and after well-known
operations extract required sum rules.

Let us begin from the phenomenological side of the sum rules, i. e. from
function $\Pi _{\mu \nu }^{\mathrm{Phys}}(p)$. We assume that the tetraquark
$Y(4660)$ with the chosen quark content and diquark-antidiquark structure is
the ground-state particle in its class. Then by introducing into Eq.\ (\ref%
{eq:CF1}) the full set of corresponding states, performing the integration
over $x$ and isolating contribution to $\Pi _{\mu \nu }^{\mathrm{Phys}}(p)$
of the ground-state, we obtain [for brevity, in formulas, we use $Y\equiv
Y(4660)$]
\begin{equation}
\Pi _{\mu \nu }^{\mathrm{Phys}}(p)=\frac{\langle 0|J_{\mu }|Y(p)\rangle
\langle Y(p)|J_{\nu }^{\dagger }|0\rangle }{m_{Y}^{2}-p^{2}}+\ldots ,
\label{eq:Phys1}
\end{equation}%
where $m_{Y}$ is the mass of $Y(4660)$ and dots show the contribution of the
higher resonances and continuum. We simplify this formula by introducing the
matrix element%
\begin{equation}
\langle 0|J_{\mu }|Y(p)\rangle =m_{Y}f_{Y}\varepsilon _{\mu }
\label{eq:Mel1}
\end{equation}%
with $f_{Y}$ and $\varepsilon _{\mu }$ being the coupling and polarization
vector of the resonance $Y(4660)$, respectively. After some simple
calculations we get%
\begin{equation}
\Pi _{\mu \nu }^{\mathrm{Phys}}(p)=\frac{m_{Y}^{2}f_{Y}^{2}}{m_{Y}^{2}-p^{2}}%
\left( -g_{\mu \nu }+\frac{p_{\mu }p_{\nu }}{p^{2}}\right) +\ldots .
\label{eq:Phys2}
\end{equation}%
It is evident that $\Pi _{\mathrm{V}}^{\mathrm{Phys}}(p^{2})=$ $%
m_{Y}^{2}f_{Y}^{2}/(m_{Y}^{2}-p^{2})$ is the invariant amplitude that can be
used later to derive sum rules.

To find $\Pi _{\mu \nu }^{\mathrm{OPE}}(p)$ we follow the recipes that have
just been outlined above and express it in terms of the quark propagators
\begin{eqnarray}
&&\Pi _{\mu \nu }^{\mathrm{OPE}}(p)=i\int d^{4}xe^{ipx}\epsilon \widetilde{%
\epsilon }\epsilon ^{\prime }\widetilde{\epsilon }^{\prime }\left\{ \mathrm{%
Tr}\left[ \gamma _{5}\widetilde{S}_{s}^{bb^{\prime }}(x)\gamma
_{5}S_{c}^{cc^{\prime }}(x)\right] \right.   \notag \\
&&\times \mathrm{Tr}\left[ \gamma _{5}\gamma _{\mu }\widetilde{S}%
_{c}^{e^{\prime }e}(-x)\gamma _{\nu }\gamma _{5}S_{s}^{d^{\prime }d}(-x)%
\right] +\mathrm{Tr}\left[ \gamma _{5}\gamma _{\mu }\widetilde{S}%
_{c}^{ee^{\prime }}(-x)\gamma _{5}\right.   \notag \\
&&\times \left. S_{s}^{d^{\prime }d}(-x)\right] \mathrm{Tr}\left[ \gamma
_{5}\gamma _{\nu }\widetilde{S}_{s}^{bb^{\prime }}(x)\gamma
_{5}S_{c}^{cc^{\prime }}(x)\right] +\mathrm{Tr}\left[ \gamma _{5}\widetilde{S%
}_{c}^{ee^{\prime }}(-x)\right.   \notag \\
&&\times \left. \gamma _{\nu }\gamma _{5}S_{s}^{d^{\prime }d}(-x)\right]
\mathrm{Tr}\left[ \gamma _{5}\widetilde{S}_{s}^{bb^{\prime }}(x)\gamma _{\mu
}\gamma _{5}S_{c}^{cc^{\prime }}(x)\right] +\mathrm{Tr}\left[ \gamma
_{5}\gamma _{\nu }\right.   \notag \\
&&\left. \times \left. \widetilde{S}_{s}^{bb^{\prime }}(x)\gamma _{\mu
}\gamma _{5}S_{c}^{cc^{\prime }}(x)\right] \mathrm{Tr}\left[ \gamma _{5}%
\widetilde{S}_{c}^{ee^{\prime }}(-x)\gamma _{5}S_{s}^{d^{\prime }d}(-x\right]
\right\} ,  \notag \\
&&  \label{eq:OPE1}
\end{eqnarray}%
where
\begin{equation*}
\widetilde{S}_{c(s)}(x)=CS_{c(s)}^{\mathrm{T}}(x)C,
\end{equation*}%
and $S_{c(s)}(x)$ is the heavy $c$-quark (the light $s$-quark) propagator.

The expressions of the quark propagators are well known and therefore we do
not provide them here explicitly (see, for example, Appendix in Ref.\ \cite%
{Agaev:2017tzv}). We calculate $\Pi _{\mu \nu }^{\mathrm{OPE}}(p)$ by taking
into account various vacuum condensates up to dimension 10 and write the QCD
counterpart of the phenomenological function $\Pi _{\mathrm{V}}^{\mathrm{OPE}%
}(p^{2})$ in terms of the corresponding spectral density $\rho (s)$%
\begin{equation}
\Pi _{\mathrm{V}}^{\mathrm{OPE}}(p^{2})=\int_{{\mathcal{M}^{2}}}^{\infty }%
\frac{\rho (s)}{s-p^{2}}ds,  \label{eq:OPE2}
\end{equation}%
where $\mathcal{M}^{2}=4(m_{c}+m_{s})^{2}$. Now, to extract the required sum
rules, we equate these invariant amplitudes to each other, apply the Borel
transformation to both sides of the obtained expression to suppress
contributions arising from the higher resonances and continuum, and perform
the continuum subtraction by utilizing the assumption about the quark-hadron
quality. The second equality can be derived by acting to the first
expression by the operator $d/d(-1/M^{2})$; these two equalities can be used
to extract the sum rules for $m_{Y}$ and $f_{Y}$:

\begin{equation}
m_{Y}^{2}=\frac{\int_{\mathcal{M}^{2}}^{s_{0}}ds\rho (s)se^{-s/M^{2}}}{\int_{%
\mathcal{M}^{2}}^{s_{0}}ds\rho (s)e^{-s/M^{2}}}  \label{eq:SR1}
\end{equation}%
and
\begin{equation}
f_{Y}^{2}=\frac{1}{m_{Y}^{2}}\int_{\mathcal{M}^{2}}^{s_{0}}ds\rho
(s)e^{(m_{Y}^{2}-s)/M^{2}}.  \label{eq:SR2}
\end{equation}%
In the sum rules given by Eqs.\ (\ref{eq:SR1}) and (\ref{eq:SR2}) $M^{2}$ is
the Borel parameter that has been introduced when applying the corresponding
transformation, and $s_{0}$ is the continuum threshold parameter that
separates the ground-state contribution from other effects.

Apart from the auxiliary parameters $M^{2}$ and $s_{0}$, the sum rules
depend also on the numerous vacuum condensates. In numerical computations,
we use their values fixed at the normalization scale $\mu _{0}^{2}=1\
\mathrm{GeV}^{2}$: for the quark and mixed condensates, $\langle \bar{q}%
q\rangle =-(0.24\pm 0.01)^{3}\ \mathrm{GeV}^{3}$, $\ \langle \bar{s}s\rangle
=0.8\ \langle \bar{q}q\rangle $, $m_{0}^{2}=(0.8\pm 0.1)\ \mathrm{GeV}^{2}$,
$\ \langle \overline{q}g_{s}\sigma Gq\rangle =m_{0}^{2}\langle \overline{q}%
q\rangle $, and $\langle \overline{s}g_{s}\sigma Gs\rangle =m_{0}^{2}\langle
\bar{s}s\rangle $, and for the gluon condensates $\langle \alpha
_{s}G^{2}/\pi \rangle =(0.012\pm 0.004)\,\mathrm{GeV}^{4}$ and $\langle
g_{s}^{3}G^{3}\rangle =(0.57\pm 0.29)\ \mathrm{GeV}^{6}$. For the masses of
the quarks we employ $m_{s}=(128\pm 10)~\mathrm{MeV}$ and $m_{c}=(1.27\pm
0.03)~\mathrm{GeV}$ borrowed from Ref.\ \cite{Patrignani:2016xqp}.

The vacuum condensates have fixed numerical values, whereas the Borel and
continuum threshold parameters can be varied within some regions, which have
to satisfy the standard restrictions of the sum rules computations. Thus,
the window for $M^{2}\in \lbrack M_{\mathrm{max}}^{2},\ M_{\mathrm{min}}^{2}]
$ is fixed from the constraints imposed on the pole contribution ($\mathrm{PC%
}$)
\begin{equation}
\mathrm{PC}=\frac{\Pi _{\mathrm{V}}(M_{\mathrm{max}}^{2},\ s_{0})}{\Pi _{%
\mathrm{V}}(M_{\mathrm{max}}^{2},\ \infty )}\geq 0.15  \label{eq:Cond1}
\end{equation}%
which determines $M_{\mathrm{max}}^{2},$ and on the ratio $R(M_{\mathrm{min}%
}^{2})$
\begin{equation}
R(M_{\mathrm{min}}^{2})=\frac{\Pi _{\mathrm{V}}^{\mathrm{DimN}}(M_{\mathrm{%
min}}^{2},\ s_{0})}{\Pi _{\mathrm{V}}(M_{\mathrm{min}}^{2},\ s_{0})}<0.05
\label{eq:Cond2}
\end{equation}%
necessary to find $M_{\mathrm{min}}^{2}$. In the expressions above \ $\Pi _{%
\mathrm{V}}(M^{2},s_{0})$ is the Borel transformed and subtracted expression
of the invariant function $\Pi _{\mathrm{V}}^{\mathrm{OPE}}(p^{2})$, and $M_{%
\mathrm{max}}^{2}$ and $M_{\mathrm{min}}^{2}$ are the maximal and minimal
allowed values of the Borel parameter. In Eq.\ (\ref{eq:Cond2}) $\Pi _{%
\mathrm{V}}^{\mathrm{DimN}}(M_{\mathrm{min}}^{2},\ \infty )$ is the
contribution to the correlation function of the last $\mathrm{Nth}$ term (or
a sum of  the last few terms) in the operator product expansion (OPE). The
ratio $R(M_{\mathrm{min}}^{2})$ quantifies the convergence of the OPE and
will be used for the numerical analysis. The last restriction on the lower
limit $M_{\mathrm{min}}^{2}$ is the prevalence of the perturbative
contribution over the nonperturbative one.

It is clear that $m_{Y}$ and $f_{Y}$ should not depend on the auxiliary
parameters $M^{2}$ and $s_{0}$. But in real calculations, these quantities
are nevertheless  sensitive to the choice  of both $M^{2}$ and $s_{0}$.
Therefore, the parameters $M^{2}$ and $s_{0}$ should  also be determined in
such a way as to minimize the dependence of $m_{Y}$ and $f_{Y}$ on them.

The analysis carried out by taking into account all of the aforementioned
constraints allows us to determine
\begin{equation}
M^{2}\in \lbrack 4.9,\ 6.8]\ \mathrm{GeV}^{2},\ s_{0}\in \lbrack 23.2,\
25.2]\ \mathrm{GeV}^{2},  \label{eq:Wind}
\end{equation}%
as the optimal regions for $M^{2}$ and $s_{0}$. In fact, at $M_{\mathrm{min}%
}^{2}$ the convergence of the operator product expansion is fulfilled with
high accuracy and $R(4.8\ \mathrm{GeV}^{2})=0.017$, which is estimated by
employing the sum of the last three terms, i.e., $\mathrm{DimN\equiv
Dim8+Dim9+Dim10}$. Moreover, at $M_{\mathrm{min}}^{2}$ the perturbative
contribution amounts to more than $74\%$ of the full result, considerably
overshooting the nonperturbative effects. The pole contribution is $\mathrm{%
PC}=0.16$, which is typical for sum rules involving multiquark aggregations.
It is worth noting that $\mathrm{PC}$ at $M_{\mathrm{min}}^{2}$ reaches its
maximal value and becomes equal to $0.78$.

\begin{widetext}

\begin{figure}[h!]
\begin{center} \includegraphics[%
totalheight=6cm,width=8cm]{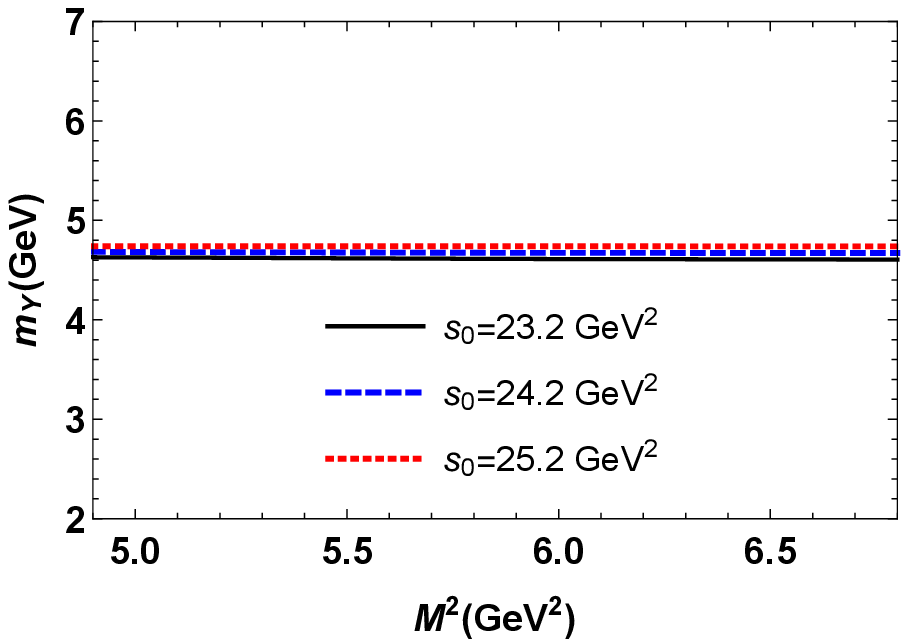}\,\,
\includegraphics[
totalheight=6cm,width=8cm]{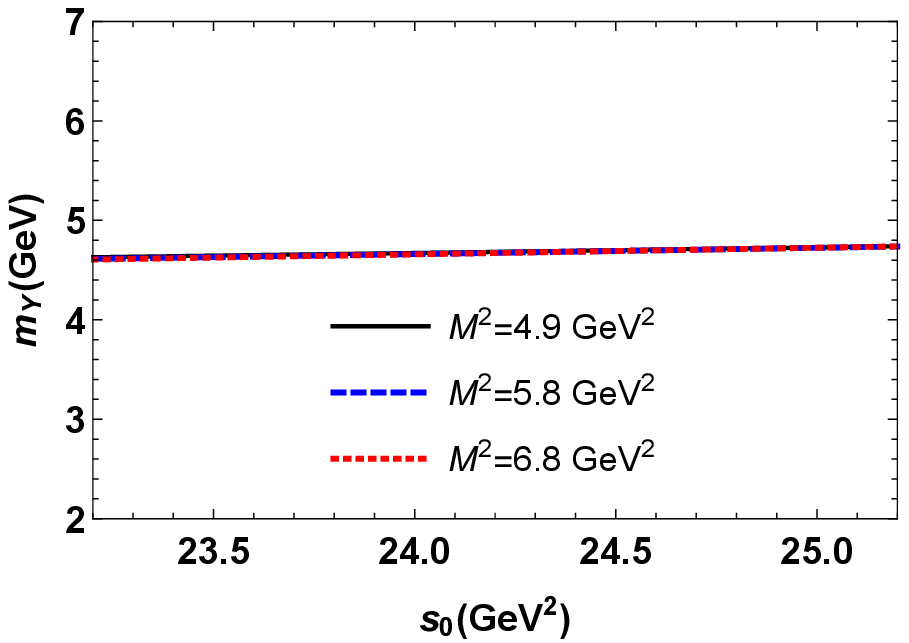}
\end{center}
\caption{ The dependence of the $Y(4660)$ resonance's mass on the Borel (left) and continuum threshold
(right) parameters.}
\label{fig:Mass}
\end{figure}
\begin{figure}[h!]
\begin{center}
\includegraphics[%
totalheight=6cm,width=8cm]{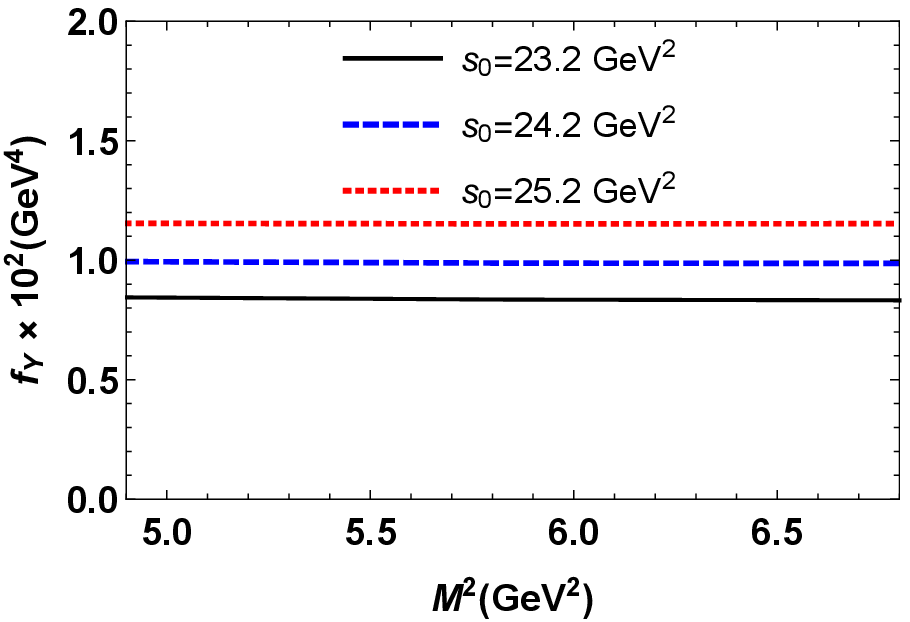}\,\,
\includegraphics[
totalheight=6cm,width=8cm]{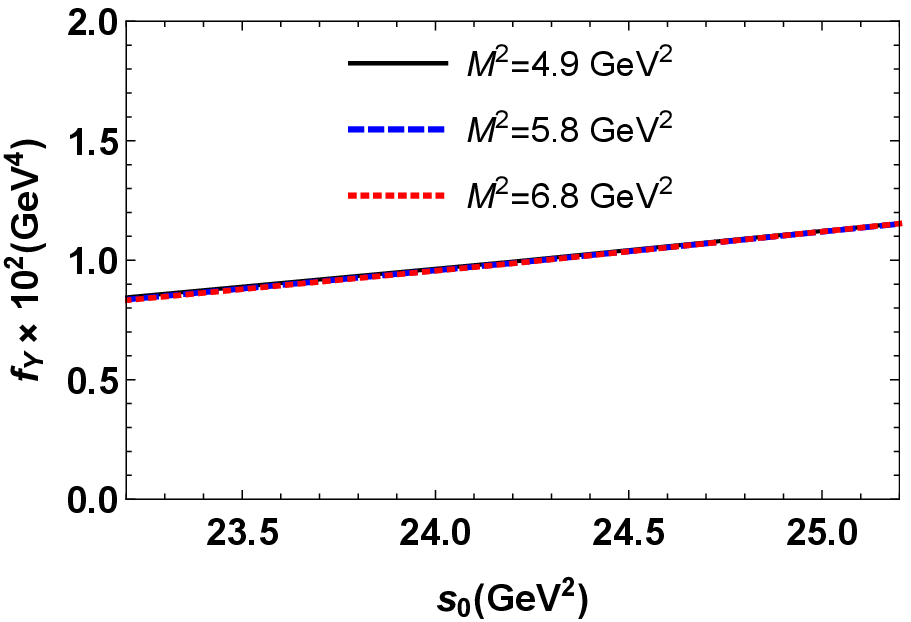}
\end{center}
\caption{ The same as in Fig.\ \ref{fig:Mass} but for the coupling $f_{Y}$.}
\label{fig:Coupling}
\end{figure}

\end{widetext}

In Figs.\ \ref{fig:Mass} and \ref{fig:Coupling} we plot the predictions for $%
m_{Y}$ and $f_{Y}$, which visually demonstrate their dependence on the used
values of $M^{2}$ and $s_{0}$. It is seen that the dependence of the mass
and coupling on the Borel parameter is very weak: the predictions for $m_{Y}$
and $f_{Y}$ demonstrate a high stability against changes of $M^{2}$ inside
of the optimized working interval. But $m_{Y}$ and $f_{Y}$ are sensitive to
the choice of the continuum threshold parameter $s_{0}$. Namely this
dependence generates a main part of uncertainties in the present sum rules,
which, nevertheless,  remain within standard limits accepted for such a kind
of computations. From these studies we extract the mass and coupling of the
resonance $Y(4660)$ as
\begin{eqnarray}
m_{Y} &=&4677_{-63}^{+71}~\mathrm{MeV},  \notag \\
f_{Y} &=&(0.99\pm 0.16)\cdot 10^{-2}\ \mathrm{GeV}^{4}.  \label{eq:CMass}
\end{eqnarray}

Our result for $m_{Y}$ is in  reasonable agreement with experimental data
\cite{Wang:2014hta}. It is also instructive to compare $m_{Y}$ with results
of other theoretical studies. As we have mentioned above, in the context of
the sum rule method the mass of the resonance $Y(4660)$ was evaluated in
different papers. Thus, in Ref.\ \cite{Zhang:2010mw}, the mass of $Y(4660)$
was found equal to $m_{Y}=(4.69\pm 0.36)\ \mathrm{GeV}$, where the authors
examined it as $P$-wave excitation of the scalar tetraquark $[cs][\overline{c%
}\overline{s}]$.

The resonance $Y(4660)$ was treated as the vector $[cs][\overline{c}%
\overline{s}]$ tetraquark in Ref.\ \cite{Albuquerque:2008up}, the mass of
which was found equal to
\begin{equation}
m_{Y}=(4.65\pm 0.10)\ \mathrm{GeV}.
\end{equation}%
These predictions are compatible with experimental data, and, by taking into
account the theoretical errors, also with our result.

The vector tetraquarks with positive and negative $C$-parities were explored
in Ref.\ \cite{Chen:2010ze}, and their masses were extracted from two-point
sum rules by taking into account vacuum condensates up to dimension 8. The
resonance $Y(4660)$ was identified in Ref.\ \cite{Chen:2010ze} as the
tetraquarks with $J^{PC}=1^{--}$ and $[cq][\overline{c}\overline{q}]$ or $%
[cs][\overline{c}\overline{s}]$ contents.

In the case of the $[cs][\overline{c}\overline{s}]$ state built of the
scalar diquark and vector antidiquark the authors used two interpolating
currents denoted in Ref. \cite{Chen:2010ze} as $J_{1\mu }$ and $J_{3\mu }$,
the first of which overshoots the mass of the $Y(4660)$ resonance
\begin{equation}
m_{J_{1}}=(4.92\pm 0.10)\ \mathrm{GeV},
\end{equation}%
whereas the second one underestimates it, leading to the result
\begin{equation}
m_{J_{3}}=(4.52\pm 0.10)\ \mathrm{GeV}.
\end{equation}%
These predictions contradict to the experimental data and also do not
coincide with our present result not the result of Ref.\ \cite%
{Albuquerque:2008up} obtained using the current Eq.\ (\ref{eq:Curr1}).

The $Y(4660)$ was assigned in Ref.\ \cite{Wang:2018rfw} to be the $C\otimes
\gamma _{\mu }C$-type vector tetraquark with the mass $m_{Y}=(4.66\pm 0.09)\
\mathrm{GeV}$ and the pole residue $\lambda _{Y}=(6.74\pm 0.88)\cdot
10^{-2}\ \mathrm{GeV}^{5}\ $, which for the coupling $f_{Y}$ leads to $%
f_{Y}=(1.45\pm 0.19)\cdot 10^{-2}\ \mathrm{GeV}^{4}$. The discrepancy
between this prediction and our result (\ref{eq:CMass}) for $f_{Y}$ can be
explained by the different assumptions on the internal structure of the
vector resonance $Y(4660)$. Indeed, in the present work we consider it  a
state composed of a scalar diquark and vector antidiquark, whereas in Ref. \
\cite{Wang:2018rfw}, it was treated as a bound state of a pseudoscalar
diquark and axial-vector antidiquark.

As is seen, within the sum rule method the $Y(4660)$ resonance can be
interpreted as the vector tetraquark $[cs][\overline{c}\overline{s}]$, but
with the different internal structures and interpolating currents.
Therefore, one has to deepen the analysis and consider decays of the state $%
Y(4660)$ to make a choice between existing models. In the next section we
are going to concentrate on the strong decay modes of $Y(4660$, in which our
results for $m_{Y}$ and $f_{Y}$ will be used as the input parameters.


\section{Strong decays of the resonance $Y(4660)$}

\label{sec:Decays}

The strong decays of the tetraquark $Y(4660)$ can be fixed using the
kinematical restriction which is evident from Eq.\ (\ref{eq:CMass}). Because
we are interested in $S$-wave decays of $Y(4660)$ the spin in these
processes should be conserved. Another constraint on possible partial decay
modes of the $Y(4660)$ tetraquark is imposed by $P$-parities of the final
particles. Performed analysis allows us to see that partial decays to $%
J/\psi f_{0}(980),$ $\psi (2S)f_{0}(980)$ and $J/\psi f_{0}(500),$\ $\psi
(2S)f_{0}(500)$ are among important decay modes of $Y(4660)$.

The $Y(4660)$ resonance's decays contain in the final state the scalar
mesons $f_{0}(980)$ and $f_{0}(500)$, which we are going to treat as
diquark-antidiquark states. The interpretation of the mesons belonging to
the light scalar nonet as four-quark systems is not new and starts from the
analyses of Refs.\ \cite{Jaffe:1976ig,Weinstein:1982gc}. In the model
suggested recently in Ref.\ \cite{Kim:2017yvd} the isoscalar mesons $%
f_{0}(980)$ and $f_{0}(500)$ are considered as mixtures of the basic
tetraquark states $\mathbf{L}=[ud][\overline{u}\overline{d}]$ and $\mathbf{H}%
=([su][\overline{s}\overline{u}]+[ds][\overline{d}s])/\sqrt{2}$. $\mathbf{\ }
$Calculations performed using this new model led to reasonable predictions
for the mass and full width of the mesons $f_{0}(980)$ and $f_{0}(500)$ \cite%
{Agaev:2017cfz,Agaev:2018sco}; these will be used in the present work, as
well. It is worth noting that this mixing phenomenon allows one to study the
decays of the $Y(4660)$ resonance to the $f_{0}(980)$ and $f_{0}(500)$
mesons within the same framework, because both of them interact with $Y(4660)
$ through their $\mathbf{H}$ components.

We concentrate on the decays $J/\psi f_{0}(980)$ and $\psi (2S)f_{0}(980)$
and calculate the strong couplings $g_{YJf_{0}(980)}$ and $g_{Y\Psi
f_{0}(980)}$ corresponding to the vertices $YJ/\psi f_{0}(980)$ and $Y\psi
(2S)f_{0}(980)$, respectively. For these purposes we employ the LCSR method
and consider the correlation function
\begin{equation}
\Pi _{\mu \nu }(p,q)=i\int d^{4}xe^{ipx}\langle f_{0}(q)|\mathcal{T}\{J_{\mu
}^{\psi }(x)J_{\nu }^{\dagger }(0)\}|0\rangle ,  \label{eq:CF2}
\end{equation}%
where $J_{\nu }(x)$ and $J_{\mu }^{\psi }(x)$ are the interpolating currents
to $Y(4660)$ and $J/\psi $, respectively. The current $J_{\nu }(x)$ has been
defined in Eq.\ (\ref{eq:Curr1}), whereas $J_{\mu }^{\psi }(x)$ is given by
the expression
\begin{equation}
J_{\mu }^{\psi }(x)=\overline{c}_{i}(x)i\gamma _{\mu }c_{i}(x).
\label{eq:Curr2}
\end{equation}%
In the vertices $p$, $q$ and $p^{\prime }=p+q$ are the momenta of $J/\psi $
or $\psi (2S)$, $f_{0}(980)$ and $Y(4660)$, respectively.

To derive the sum rules for $g_{YJf_{0}(980)}$ and $g_{Y\Psi f_{0}(980)}$ we
first calculate $\Pi _{\mu \nu }(p,q)$ in terms of the physical parameters
of involved particles. \ It is not difficult to get
\begin{eqnarray}
&&\Pi _{\mu \nu }^{\mathrm{Phys}}(p,q)=\frac{\langle 0|J_{\mu }^{\psi
}|J/\psi \left( p\right) \rangle }{p^{2}-m_{J}^{2}}\langle J/\psi \left(
p\right) f_{0}(q)|Y(p^{\prime })\rangle   \notag \\
&&\times \frac{\langle Y(p^{\prime })|J_{\nu }^{\dagger }|0\rangle }{%
p^{\prime 2}-m_{Y}^{2}}+\frac{\langle 0|J_{\mu }^{\psi }|\psi (2S)\left(
p\right) \rangle }{p^{2}-m_{\psi }^{2}}  \notag \\
&&\times \langle \psi (2S)\left( p\right) f_{0}(q)|Y(p^{\prime })\rangle
\frac{\langle Y(p^{\prime })|J_{\nu }^{\dagger }|0\rangle }{p^{\prime
2}-m_{Y}^{2}}\ldots ,  \label{eq:CF3}
\end{eqnarray}%
where $m_{J}$ and $m_{\psi }$ are the masses of the mesons $J/\psi $ and $%
\psi (2S)$, respectively. The dots in Eq.\ (\ref{eq:CF3}) denote a
contribution of the higher resonances and continuum states. As is seen, $\Pi
_{\mu \nu }^{\mathrm{Phys}}(p,q)$ contains two terms and corresponds to the
"ground-state+first radially excited state + continuum " scheme.

Further simplification of $\Pi _{\mu \nu }^{\mathrm{Phys}}(p,q)$ can be
achieved by employing the matrix element (\ref{eq:Mel1}) and new ones from
Eq.\ (\ref{eq:Mel2})
\begin{eqnarray}
&&\langle 0|J_{\mu }^{\psi }|J/\psi \left( p\right) \rangle
=f_{J}m_{J}\varepsilon _{\mu },  \notag \\
&&\langle 0|J_{\mu }^{\psi }|\psi (2S)\left( p\right) \rangle =f_{\psi
}m_{\psi }\varepsilon _{\mu },  \label{eq:Mel2}
\end{eqnarray}%
as well as by introducing two elements that describe the vertices
\begin{eqnarray}
&&\langle J/\psi \left( p\right) f_{0}(q)|Y(p^{\prime })\rangle
=g_{YJf_{0}(980)}\left[ (p\cdot p^{\prime })\right.   \notag \\
&&\left. \times (\varepsilon ^{\ast }\cdot \varepsilon ^{\prime })-(p\cdot
\varepsilon ^{\prime })(p^{\prime }\cdot \varepsilon ^{\ast })\right] ,
\notag \\
&&\langle \psi (2S)\left( p\right) f_{0}(q)|Y(p^{\prime })\rangle =g_{Y\Psi
f_{0}(980)}\left[ (p\cdot p^{\prime })\right.   \notag \\
&&\left. \times (\varepsilon ^{\ast }\cdot \varepsilon ^{\prime })-(p\cdot
\varepsilon ^{\prime })(p^{\prime }\cdot \varepsilon ^{\ast })\right] .
\label{eq:Mel3}
\end{eqnarray}%
In the expressions above,  $f_{J}$ ($f_{\psi }$) is the $J/\psi $ [$\psi (2S)
$] meson's decay constant, and $\varepsilon _{\mu }$ and $\varepsilon _{\nu
}^{\prime }$ are the polarization vectors of the $J/\psi $ [$\psi (2S)$]
meson and the resonance $Y(4660)$, respectively.

Then the correlation function takes the following form
\begin{eqnarray}
&&\Pi _{\mu \nu }^{\mathrm{Phys}}(p,q)=\frac{%
g_{YJf_{0}(980)}f_{J}m_{J}f_{Y}m_{Y}}{\left( p^{\prime 2}-m_{Y}^{2}\right)
\left( p^{2}-m_{J}^{2}\right) }\left( -p_{\mu }^{\prime }p_{\nu }\right.
\notag \\
&&\left. +\frac{m_{Y}^{2}+m_{J}^{2}}{2}g_{\mu \nu }\right) +\frac{g_{Y\Psi
f_{0}(980)}f_{\psi }m_{\psi }f_{Y}m_{Y}}{\left( p^{\prime
2}-m_{Y}^{2}\right) \left( p^{2}-m_{\psi }^{2}\right) }  \notag \\
&&\times \left( -p_{\mu }^{\prime }p_{\nu }+\frac{m_{Y}^{2}+m_{\psi }^{2}}{2}%
g_{\mu \nu }\right) +\ldots .  \label{eq:CF4}
\end{eqnarray}%
We extract the sum rules for the strong couplings using the invariant
functions corresponding to the structure $\sim g_{\mu \nu }$. The
correlation function $\Pi _{\mu \nu }(p,q)$ contains inside of the $\mathcal{%
T}$-operation a tetraquark and  conventional meson currents, therefore, this
situation does not differ considerably from the analysis of the
tetraquark-meson-meson vertices elaborated in Ref.\ \cite{Agaev:2016dev}.
These vertices can be investigated using the $q\rightarrow 0$ limit of the
full LCSR method, which is known as the "soft-meson approximation" \cite%
{Belyaev:1994zk,Ioffe:1983ju}. This approximation was applied numerously to
study decays of the tetraquarks, for example, in Refs.\ \cite%
{Agaev:2016ijz,Agaev:2016dsg,Agaev:2017foq}.

 In the general case the invariant function $\Pi ^{\mathrm{Phys}%
}(p^{2},p^{\prime 2})$ depends on two variables, but in the soft
approximation when $p=p^{\prime }$ it reduces to $\Pi ^{\mathrm{Phys}}(p^{2})
$. \ In this approach we replace $1/[\left( p^{\prime 2}-m_{Y}^{2}\right)
\left( p^{2}-m_{J}^{2}\right) ]$ by the double pole factor $1/[\left(
p^{2}-m_{1}^{2}\right) ^{2}]$, where $m_{1}^{2}=(m_{Y}^{2}+m_{J}^{2})/2$.
The same is true also for the second term in Eq.\ (\ref{eq:CF4}) with the
clear replacement $m_{1}^{2}\rightarrow m_{2}^{2}=(m_{Y}^{2}+m_{\psi }^{2})/2
$. Then, the Borel transformation of the $\Pi ^{\mathrm{Phys}}(p^{2})$ reads
\begin{eqnarray}
&&\mathcal{B}\Pi ^{\mathrm{Phys}%
}(p^{2})=g_{YJf_{0}(980)}f_{J}m_{J}f_{Y}m_{Y}m_{1}^{2}\frac{%
e^{-m_{1}^{2}/M^{2}}}{M^{2}}  \notag \\
&&+g_{Y\Psi f_{0}(980)}f_{\psi }m_{\psi }f_{Y}m_{Y}m_{2}^{2}\frac{%
e^{-m_{2}^{2}/M^{2}}}{M^{2}}\ldots .  \label{eq:CF5}
\end{eqnarray}

It the next step, one has to find the expression of the correlation function
in terms of the quark propagators. After some calculations, we get
\begin{eqnarray}
&&\Pi _{\mu \nu }^{\mathrm{OPE}}(p,q)=\int d^{4}xe^{ipx}\epsilon \widetilde{%
\epsilon }\left[ \gamma _{5}\widetilde{S}_{c}^{ic}(x){}\gamma _{\mu }\right.
\notag \\
&&\left. \times \widetilde{S}_{c}^{ei}(-x){}\gamma _{\nu }\gamma _{5}-\gamma
_{\nu }\gamma _{5}\widetilde{S}_{c}^{ic}(x){}\gamma _{\mu }\widetilde{S}%
_{c}^{ei}(-x){}\gamma _{5}\right] _{\alpha \beta }  \notag \\
&&\times \langle f_{0}(q)|\overline{s}_{\alpha }^{b}(0)s_{\beta
}^{d}(0)|0\rangle ,  \label{eq:CF6}
\end{eqnarray}%
where $\alpha $ and $\beta $ are the spinor indices.

The matrix element $\langle f_{0}(q)|\overline{s}_{\alpha }^{b}(0)s_{\beta
}^{d}(0)|0\rangle $ has to be rewritten in a form suitable for further
analysis. To this end, we apply the expansion
\begin{equation}
\overline{s}_{\alpha }^{b}s_{\beta }^{d}\rightarrow \frac{1}{12}\delta
_{bd}\Gamma _{\beta \alpha }^{j}\left( \overline{s}\Gamma ^{j}s\right) ,
\label{eq:MatEx}
\end{equation}%
where $\Gamma ^{j}=\mathbf{1,\ }\gamma _{5},\ \gamma _{\lambda },\ i\gamma
_{5}\gamma _{\lambda },\ \sigma _{\lambda \rho }/\sqrt{2}$ form the full set
of Dirac matrices, and express $\Pi _{\mu \nu }^{\mathrm{OPE}}(p,q)$ in
terms of the local matrix elements of the scalar meson $f_{0}(980)$. \
Calculations prove that the matrix elements with $\Gamma ^{j}=\mathbf{\ }%
\gamma _{5}$ and $\ i\gamma _{5}\gamma _{\lambda },$ i.e., ones with an odd
number of $\gamma _{5}$ matrices are identically equal to zero. The matrix
elements in Eq.\ (\ref{eq:MatEx}) with $\gamma _{\lambda }$ and $\sigma
_{\lambda \rho }/\sqrt{2}$ should be proportional to $q_{\lambda }$ and $%
q_{\lambda }q_{\rho \text{ }}$because only the momentum of $f_{0}(980)$ has
the required Lorentz index. But in the soft approximation, $q=0$, and
therefore these elements do not contribute to $\Pi _{\mu \nu }^{\mathrm{OPE}%
}(p,q)$. In the matrix element with $\ \sigma _{\lambda \rho }/\sqrt{2}$
components, $\sim g_{\lambda \rho }$ may lead to some effects, but in the
present work we neglect them. \ We also ignore matrix elements $\sim G$ with
insertions of the gluon field strength tensor, contributions of which in the
soft approximation, as a rule, vanish. Hence, the only matrix element that
we take into account is
\begin{equation}
\langle f_{0}(980)(q)|\overline{s}(0)s(0)|0\rangle =\lambda _{f^{\prime }},
\label{eq:MatE2}
\end{equation}%
which forms the correlation function $\Pi _{\mu \nu }^{\mathrm{OPE}}(p,q=0)$%
. \ The $\lambda _{f^{\prime }}$ and the similar matrix element $\langle
f_{0}(500)(q)|\overline{s}(0)s(0)|0\rangle =\lambda _{f}$ can be computed
using the two-point sum rule method, the details of which are presented in
the Appendix.

After standard calculations for the Borel transformed correlation function $%
\Pi ^{\mathrm{OPE}}(M^{2})$ we find%
\begin{eqnarray}
&&\Pi ^{\mathrm{OPE}}(M^{2})=\frac{\lambda _{f^{\prime }}}{24\pi ^{2}}%
\int_{4m_{c}^{2}}^{\infty }\frac{ds}{s}\sqrt{s(s-4m_{c}^{2})}  \notag \\
&&\times (s+8m_{c}^{2})+\lambda _{f^{\prime
}}\int_{0}^{1}dze^{-m_{c}^{2}/M^{2}Z}F(z,M^{2}),  \label{eq:CF10}
\end{eqnarray}%
where the first term is the perturbative contribution, whereas the
nonperturbative effects are encoded by the second term. The function $%
F(z,M^{2})$ in Eq.\ (\ref{eq:CF10}) has the following form%
\begin{eqnarray}
&&F(z,M^{2})=-\frac{\left\langle \alpha _{s}G^{2}/\pi \right\rangle m_{c}^{2}%
}{72M^{4}}\frac{1}{Z}\left[ m_{c}^{2}\left( 1-2Z\right) \right.  \notag \\
&&\left. -M^{2}Z\left( 3-7Z\right) \right] +\frac{\langle
g_{s}^{3}G^{3}\rangle }{45\cdot 2^{9}\pi ^{2}M^{8}Z^{5}}  \notag \\
&&\times \left\{ m_{c}^{6}(1-2z)^{2}(9-11Z)+2m_{c}^{2}M^{4}Z^{2}\right.
\notag \\
&&\times \lbrack -42+Z(122+9Z)]-2M^{6}Z^{3}  \notag \\
&&\times \left[ 6-Z(22-9Z)\right] +m_{c}^{4}M^{2}Z\left( -11+119Z\right.
\notag \\
&&\left. \left. -190Z^{2}\right) \right\} +\frac{\left\langle \alpha
_{s}G^{2}/\pi \right\rangle ^{2}m_{c}^{2}\pi ^{2}}{648M^{10}Z^{3}}\left[
m_{c}^{4}-m_{c}^{2}M^{2}\right.  \notag \\
&&\left. \times (1+4Z)+2M^{4}Z(2-Z)\right] ,  \label{eq:Func}
\end{eqnarray}%
where $Z=z(1-z)$.

The pertubative term in Eq.\ (\ref{eq:CF10}) is calculated as an imaginary
part of the relevant component in $\Pi _{\mu \nu }^{\mathrm{OPE}}(p,q=0)$,
and afterward, the Borel transformation are carried out. The Borel
transformation of the nonpertiurbative contribution is computed directly
from $\Pi _{\mu \nu }^{\mathrm{OPE}}(p,q=0)$ and contains vacuum condensates
up to dimension 8. By equating $\mathcal{B}\Pi ^{\mathrm{Phys}}(p^{2})$ to $%
\Pi ^{\mathrm{OPE}}(M^{2})$ and performing the continuum subtraction we find
an expression that depends on two unknown variables $g_{YJf_{0}(980)}$ and $%
g_{Y\Psi f_{0}(980)}$. Let us note that continuum subtraction in the
perturbative part is done by $\infty \rightarrow s_{0\text{ }}$ replacement.
Because all terms in Eq.\ (\ref{eq:Func}) are proportional to inverse powers
of the Borel parameter $M^{2}$, in accordance with accepted methodology
(see, Ref.\ \cite{Belyaev:1994zk}) the nonperturbative contribution should
be left in an unsubtracted form preserving its original version. The second
equation necessary for our purposes can be derived by applying the operator $%
d/d(-1/M^{2})$ to both sides of this expression. These two equalities allow
us to find sum rules for both $g_{YJf_{0}(980)}$ and $g_{Y\Psi f_{0}(980)}$,
the explicit formulas of which are too cumbersome to present here.

The width of the decay process, for example, $Y\rightarrow \psi
(2S)f_{0}(980)$, can be found by means of the formula%
\begin{eqnarray}
&&\Gamma (Y\rightarrow \psi (2S)f_{0}(980))=\frac{g_{Y\Psi
f_{0}(980)}^{2}m_{\psi }^{2}}{24\pi }  \notag \\
&&\times \Lambda \left( 3+\frac{2\Lambda ^{2}}{m_{\psi }^{2}}\right) ,
\label{eq:DW}
\end{eqnarray}%
where $\Lambda =\Lambda (m_{Y},m_{\psi },m_{f_{0}})$ and
\begin{equation*}
\Lambda (a,b,c)=\frac{\sqrt{a^{4}+b^{4}+c^{4}-2\left(
a^{2}b^{2}+a^{2}c^{2}+b^{2}c^{2}\right) }}{2a}.
\end{equation*}

The numerical computations of the strong couplings are performed using the
values of the different vacuum condensates (see Sec. II) as well as
spectroscopic parameters of the mesons $J/\psi $ and $\psi (2S)$ (in units
of $\mathrm{MeV}$): $m_{J}=3096.900\pm 0.006$ and $f_{J}=411\pm 7$ and \ $%
m_{\psi }=3686.097\pm 0.005$ and $f_{\psi }=279\pm 8$. The parameters of the
resonance $Y(4660)$ have been found in the present work, and for the mass of
the $f_{0}(980)$ meson we use its experimentally measured value $%
m_{f_{0}}=990\pm 20$ $\mathrm{MeV}$. The parameters $M^{2}$ and $s_{0}$ are
varied inside of the regions: $M^{2}=(4.9-6.8)\ \mathrm{GeV}^{2}$ and $%
s_{0}=(23.2-25.2)\ \mathrm{GeV}^{2}$. The obtained results for the strong
couplings read%
\begin{eqnarray}
|g_{YJf_{0}(980)}| &=&(0.22\pm 0.07)~\mathrm{GeV}^{-1},  \notag \\
g_{Y\Psi f_{0}(980)} &=&(1.22\pm 0.33)~\mathrm{GeV}^{-1}.  \label{eq:SCoupl1}
\end{eqnarray}%
Then widths of the corresponding partial decay channels become equal to (in
units of $\mathrm{MeV}$):%
\begin{eqnarray}
\Gamma (Y &\rightarrow &J/\psi f_{0}(980))=18.8\pm 5.4,  \notag \\
\Gamma (Y &\rightarrow &\psi (2S)f_{0}(980))=30.2\pm 8.5.  \label{eq:DW1}
\end{eqnarray}

Analysis of the remaining two decays does not differ from previous ones and
leads to predictions
\begin{eqnarray}
g_{YJf_{0}(500)} &=&(0.07\pm 0.02)~\mathrm{GeV}^{-1},  \notag \\
|g_{Y\Psi f_{0}(500)}| &=&(0.18\pm 0.05)~\mathrm{GeV}^{-1},
\label{eq:SCoupl2}
\end{eqnarray}%
and (in $\mathrm{MeV}$)
\begin{eqnarray}
\Gamma (Y &\rightarrow &J/\psi f_{0}(500))=2.7\pm 0.7,  \notag \\
\Gamma (Y &\rightarrow &\psi (2S)f_{0}(500))=13.1\pm 3.7.  \label{eq:DW2}
\end{eqnarray}%
The total width of the $Y(4660)$ resonance estimated using these four strong
decay channels%
\begin{equation}
\Gamma _{Y}=(64.8\pm 10.8)\ \ \mathrm{MeV}  \label{eq:TWD}
\end{equation}%
is in nice agreement with the experimental value $68\pm 11\pm 1\ \mathrm{MeV}
$. For the total width of the $Y(4660)$ resonance the Particle Data Group
provides the world average $\Gamma _{Y}=72\pm 11\ \mathrm{MeV}$ \cite%
{Patrignani:2016xqp}. This is higher than the result of Ref.\ \cite%
{Wang:2014hta}, nevertheless, within uncertainties of theoretical
calculations and errors of experimental measurements the prediction obtained
here is compatible with the world average, as well. One has also to take
into account that the diquark-antidiquark model for the $Y(4660)$ implies
the existence of the $S$-wave decay channels $Y(4660)\rightarrow D_{s}^{\pm
}D_{s1}^{\mp }(2460)$ and $Y(4660)\rightarrow D_{s}^{\ast \pm }D_{s0}^{\mp
}(2317)$ that also contribute to $\Gamma _{Y}$ and may improve this
agreement.


\section{Conclusions}

\label{sec:Conc}
In the present work we have calculated the full width of the resonance $%
Y(4660)$ by interpreting it as the diquark-antidiquark state with quantum
numbers $J^{PC}=1^{--}$. Its partial decay widths depend, as important input
parameters, on the mass $m_{Y}$ and coupling $f_{Y}$. The mass of the $%
Y(4660)$ as a scalar diquark-vector antidiquark $[cs][\overline{c}\overline{s%
}]$ state was originally calculated in Refs. \cite%
{Albuquerque:2008up,Chen:2010ze}. But in these articles the coupling of the
resonance $Y(4660)$ was not evaluated. Therefore, we have computed the
spectroscopic parameters of the $Y(4660)$ state by employing the QCD
two-point sum rules and taking into account quark, gluon and mixed
condensates up to dimension 10. This has allowed us to improve the accuracy
of the aforementioned computations as well as to find the coupling of the
resonance $Y(4660)$. Our result for $m_{Y}=4677_{-63}^{+71}~\mathrm{MeV}$
within theoretical ambiguities agrees with experimental data and the
prediction made in Ref. \cite{Albuquerque:2008up} but is not compatible with
predictions of Ref. \cite{Chen:2010ze}. The coupling $f_{Y}$ in the
framework of the sum rule method was evaluated in Ref. \cite{Wang:2018rfw},
in which the another suggestion about the structure of the resonance $Y(4660)
$, namely, a pseudoscalar diquark-axial-vector antidiquark picture was
employed. The coupling $f_{Y}$ found there is larger than our result, which
can be attributed to different structures used in Ref. \cite{Wang:2018rfw}
and in the present study.

In other words, calculation of a tetraquark's mass does not provide
information enough to interpret it unambiguously as a bound state of a
diquark and an antidiquark with fixed quantum numbers. Additional important
information can be extracted from analysis of its decay channels. In the
present article, we have computed the full width of the resonance $Y(4660)$
by taking into account its $S$-wave strong decays $Y\rightarrow J/\psi
f_{0}(500)$, $Y\rightarrow \psi (2S)f_{0}(500)$, $Y\rightarrow J/\psi
f_{0}(980)$ and $Y\rightarrow \psi (2S)f_{0}(980)$ and found  reasonable
agreement with the measurements. However, the process $Y(4660)\rightarrow
\psi (2S)\pi ^{+}\pi ^{-}$ is the only decay mode of the state $Y(4660)$
observed experimentally. It is known that the dominant decay channels of $\ $%
the $f_{0}(500)$ and $f_{0}(980)$ mesons are processes $f_{0}\rightarrow \pi
^{+}\pi ^{-}$ and $f_{0}\rightarrow \pi ^{0}\pi ^{0}$. Therefore, the chains
$Y(4660)\rightarrow \psi (2S)f_{0}(980)\rightarrow \psi (2S)\pi ^{+}\pi ^{-}$
and $Y(4660)\rightarrow \psi (2S)f_{0}(500)\rightarrow \psi (2S)\pi ^{+}\pi
^{-}$ explain a dominance of the observed final state in the decay of the
resonance $Y(4660)$. In the tetraquark model, as we have seen the width of
the channel $Y(4660)\rightarrow J/\psi f_{0}(980)$ is sizeable.
Additionally, the final states $\psi (2S)\pi ^{0}\pi ^{0}$ and $J/\psi \pi
^{0}\pi ^{0}$ should also be detected. But neither $J/\psi \pi ^{+}\pi ^{-}$
nor $\pi ^{0}\pi ^{0}$ were observed in the $Y(4660)$ decays. It is worth
noting that most of the aforementioned final particles were discovered in
decays of the vector resonance $Y(4260)$: its partial decays to $J/\psi \pi
^{+}\pi ^{-}$ and $J/\psi \pi ^{0}\pi ^{0}$ as well as to $J/\psi K^{+}K^{-}$
were seen experimentally. Therefore, more accurate measurements may reveal
these modes in decays of the resonance $Y(4660)$, as well.

A situation with decays to $D_{s}$ mesons is more difficult because in the
tetraquark model there are not evident reasons for these channels of the $%
Y(4660)$ state to be highly suppressed or even forbidden. Decays to a pair
of $D$ mesons were not seen in the case of the resonance $Y(4260)$ either.
It is quite possible that partial widths of decays to $D_{s}$ mesons are
numerically small. But this is only an assumption that must be confirmed by
explicit calculations. Further experimental investigations of the $Y(4660)$
resonance, more precious measurements can enlighten problems with its decays
channels and, as a result, with its nature.


\section*{ACKNOWLEDGMENTS}

H.~S.~ and K.~A.~ thank TUBITAK for the partial financial support provided
under Grant No. 115F183.

\appendix*

\section{ The local matrix elements}

\renewcommand{\theequation}{\Alph{section}.\arabic{equation}} \label{sec:App}
In this Appendix, we calculate the couplings $\lambda _{f}$ and $\lambda
_{f^{\prime }}$ [hereafter $f=f_{0}(500)$ and $f^{\prime }=f_{0}(980)$]
defined as the matrix elements of the current $J_{_{\overline{s}s}}(x)=%
\overline{s}(x)s(x)$ sandwiched between the exotic meson and vacuum states
\begin{equation}
\langle f(q)|\overline{s}s|0\rangle =\lambda _{f},\ \ \langle f^{\prime }(q)|%
\overline{s}s|0\rangle =\lambda _{f^{\prime }}.  \label{eq:Mel4}
\end{equation}%
To this end, we explore the two-point correlation function (see, for
example, Ref.\ \cite{Zanetti:2011ju})
\begin{equation}
\Pi ^{f(f^{\prime })}(q)=i\int d^{4}xe^{iqx}\langle 0|\mathcal{T}%
\{J^{f(f^{\prime })}(x)J_{_{\overline{s}s}}^{\dagger }(0)\}|0\rangle ,
\label{eq:CF7}
\end{equation}%
where $J^{f(f^{\prime })}(x)$ is the interpolating current for the scalar
tetraquark $f$ or $f^{\prime }$. In the two-angles mixing scheme these
currents are given by the expression \cite{Agaev:2017cfz}
\begin{equation}
\begin{pmatrix}
J^{f}(x) \\
J^{f^{\prime }}(x)%
\end{pmatrix}%
=U(\varphi _{H,}\varphi _{L})%
\begin{pmatrix}
J^{H}(x) \\
J^{L}(x)%
\end{pmatrix}%
,  \label{eq:Curr3}
\end{equation}%
where $U(\varphi _{H,}\varphi _{L})$ is the mixing matrix
\begin{equation}
U(\varphi _{H,}\varphi _{L})=%
\begin{pmatrix}
\cos \varphi _{H} & -\sin \varphi _{L} \\
\sin \varphi _{H} & \cos \varphi _{L}%
\end{pmatrix}%
,
\end{equation}%
which is responsible also for the couplings' mixing.

The currents $J^{L}(x)$ and $J^{H}(x)$ correspond to the basic states $%
\mathbf{L}=[ud][\overline{u}\overline{d}]$ and $\mathbf{H}=([su][\overline{s}%
\overline{u}]+[ds][\overline{d}s])/\sqrt{2\text{ }}$ and have the following
forms%
\begin{eqnarray}
&&J^{H}(x)=\frac{\epsilon \widetilde{\epsilon }}{\sqrt{2}}\left\{ \left[
u_{a}^{T}(x)C\gamma _{5}s_{b}(x)\right] \left[ \overline{u}_{c}(x)\gamma
_{5}C\overline{s}_{e}^{T}(x)\right] \right.  \notag \\
&&\left. +\left[ d_{a}^{T}(x)C\gamma _{5}s_{b}(x)\right] \left[ \overline{d}%
_{c}(x)\gamma _{5}C\overline{s}_{e}^{T}(x)\right] \right\} ,
\end{eqnarray}%
and
\begin{equation}
J^{L}(x)=\epsilon \widetilde{\epsilon }\left[ u_{a}^{T}(x)C\gamma
_{5}d_{b}(x)\right] \left[ \overline{u}_{c}(x)\gamma _{5}C\overline{d}%
_{e}^{T}(x)\right] ,
\end{equation}%
where $\epsilon \widetilde{\epsilon }=\epsilon ^{dab}\epsilon ^{dce}$.

For an example, let us write down all expressions for the $f$ meson. To find
the phenomenological side of the sum rule, we use the
"ground-state+continuum" scheme and get
\begin{equation}
\Pi ^{f\mathrm{Phys}}(q)=\frac{\langle 0|J^{f}(x)|f(q)\rangle \langle
f(q)|J_{_{\overline{s}s}}^{\dagger }(0)|0\rangle }{m_{f}^{2}-q^{2}}+\ldots ,
\end{equation}%
where the dots traditionally stand for the higher resonances and continuum.
\ We continue using explicit expressions of the matrix elements $\langle
0|J^{f}(x)|f(q)\rangle $ and $\langle f(q)|J_{_{\overline{s}s}}^{\dagger
}(0)|0\rangle $. The former element has just been introduced by Eq.\ (\ref%
{eq:Mel4}), and after some manipulations can be recast to the final form
\begin{equation}
\langle 0|J^{f}|f(q)\rangle =m_{f}(F_{H}\cos ^{2}\varphi _{H}+F_{L}\sin
^{2}\varphi _{L}).  \label{eq:Mel5}
\end{equation}%
During this process, we have used the current $J^{f}$ as it is given in Eq. (%
\ref{eq:Curr3}) and also the matrix elements
\begin{equation}
\langle 0|J^{i}|f(p)\rangle =F_{f}^{i}m_{f},\,\ \ i=H,L.  \label{eq:Coupl}
\end{equation}%
We also benefited from the suggestion made in Ref.\ \cite{Agaev:2017cfz}
that the couplings $F_{f}^{i}$ follow a pattern of state mixing which in the
two-angles mixing scheme implies
\begin{equation}
\begin{pmatrix}
F_{f}^{H} & F_{f}^{L} \\
F_{f^{\prime }}^{H} & F_{f^{\prime }}^{L}%
\end{pmatrix}%
=U(\varphi _{H,}\varphi _{L})%
\begin{pmatrix}
F_{H} & 0 \\
0 & F_{L}%
\end{pmatrix}%
,  \label{eq:2AngleCoupl}
\end{equation}%
where $F_{H}$ and $F_{L}$ may be formally interpreted as couplings of the
"particles" $|\mathbf{H}\rangle $ and $|\mathbf{L}\rangle .$

Then we get
\begin{equation}
\Pi ^{f\mathrm{Phys}}(q)=\frac{\lambda _{f}m_{f}(F_{H}\cos ^{2}\varphi
_{H}+F_{L}\sin ^{2}\varphi _{L})}{m_{f}^{2}-q^{2}}+\ldots   \label{eq:CF8}
\end{equation}%
The following task is a computation of $\Pi ^{\mathrm{OPE}}(q)$, which leads
to
\begin{equation}
\Pi ^{f\mathrm{OPE}}(q)=\cos \varphi _{H}\Pi _{0}^{\mathrm{OPE}}(q),
\end{equation}%
where%
\begin{eqnarray}
&&\Pi _{0}^{\mathrm{OPE}}(q)=i^{2}\int d^{4}xe^{iqx}\frac{\epsilon
_{dab}\epsilon _{dae}}{6\sqrt{2}}\langle \overline{q}q\rangle   \notag \\
&&\times \mathrm{Tr}\left[ \gamma _{5}\widetilde{S}_{s}^{ie}(-x)\widetilde{S}%
_{s}^{bi}(x)\gamma _{5}\right] .  \label{eq:CF9}
\end{eqnarray}%
The matrix element $\lambda _{f}$ can be evaluated from the sum rule
\begin{equation}
\lambda _{f}=\frac{\Pi _{0}^{\mathrm{OPE}}(M^{2},\ s_{0})\cos \varphi _{H}}{%
m_{f}(F_{H}\cos ^{2}\varphi _{H}+F_{L}\sin ^{2}\varphi _{L})},
\end{equation}%
where $\Pi _{0}^{\mathrm{OPE}}(M^{2},\ s_{0})$ is the Borel transform of the
correlation function $\Pi _{0}^{\mathrm{OPE}}(q)$. The matrix element of the
$f^{\prime }$ meson can be computed by means of the same expression with
trivial replacements $m_{f}\rightarrow m_{f^{\prime }}$, $\lambda
_{f}\rightarrow \lambda _{f^{\prime }\text{ }}$, $\cos \varphi
_{H}\rightarrow \sin \varphi _{H}$ and \ $\sin \varphi _{L}\rightarrow \cos
\varphi _{L}$.

In numerical computations we have utilized the parameters of the $%
f-f^{\prime }$ system from Ref.\ \cite{Agaev:2017cfz} , i.e. for the mixing
angles we have used $\varphi _{H}=-28^{\circ }.87\pm 0^{\circ }.42$ and $%
\varphi _{L}=-\ 27^{\circ }.66\pm 0^{\circ }.31$, whereas for the couplings $%
F_{H}=(1.35\pm 0.34)\cdot 10^{-3}\ \mathrm{GeV}^{4}$ and $F_{L}=(0.68\pm
0.17)\cdot 10^{-3}\ \mathrm{GeV}^{4}$ have been employed. The masses of the
scalar particles $m_{f}=(518\pm 74)\ \mathrm{MeV}$ and $m_{f^{\prime
}}=(996\pm 130)\ \mathrm{MeV}$ have been borrowed from Ref.\ \cite%
{Agaev:2017cfz}, as well. In calculations of $\lambda _{f}$, the Borel and
continuum threshold parameters have been chosen as $M^{2}=(0.75-1.0)\
\mathrm{GeV}^{2}$ and $s_{0}=(0.8-1.1)\ \mathrm{GeV}^{2}$, whereas in the
case of $\lambda _{f^{\prime }}$ we have used $M^{2}=(1.1-1.3)\ \mathrm{GeV}%
^{2}$ and $s_{0}=(1.4-1.6)\ \mathrm{GeV}^{2}$. As a result we have found
\begin{eqnarray}
\lambda _{f} &=&(0.015\pm 0.004)\ \mathrm{GeV}^{2},  \notag \\
\ |\lambda _{f^{\prime }}| &=&(0.052\pm 0.013)\ \mathrm{GeV}^{2},
\label{eq:MEl1}
\end{eqnarray}%
which have been used in Sec. \ref{sec:Decays}.


\begin{thebibliography}{99}

\bibitem{Jaffe:1976ig} R.~L.~Jaffe,
Phys.\ Rev.\ D \textbf{15}, 267 (1977). 


\bibitem{Choi:2003ue} S.~K.~Choi \textit{et al.} [Belle Collaboration],
Phys.\ Rev.\ Lett.\ \textbf{91}, 262001 (2003).


\bibitem{Chen:2016qju} H.~X.~Chen, W.~Chen, X.~Liu and S.~L.~Zhu,
Phys.\ Rept.\ \textbf{639}, 1 (2016).


\bibitem{Esposito:2016noz} A.~Esposito, A.~Pilloni and A.~D.~Polosa,
Phys.\ Rept.\ \textbf{668}, 1 (2017).



\bibitem{Wang:2007ea} X.~L.~Wang \textit{et al.} [Belle Collaboration],
Phys.\ Rev.\ Lett.\ \textbf{99}, 142002 (2007).


\bibitem{Wang:2014hta} X.~L.~Wang \textit{et al.} [Belle Collaboration],
Phys.\ Rev.\ D \textbf{91}, 112007 (2015).


\bibitem{Pakhlova:2008vn} G.~Pakhlova \textit{et al.} [Belle Collaboration],
Phys.\ Rev.\ Lett.\ \textbf{101}, 172001 (2008).


\bibitem{Lees:2012pv} J.~P.~Lees \textit{et al.} [BaBar Collaboration],
Phys.\ Rev.\ D \textbf{89}, 111103 (2014).


\bibitem{Ding:2007rg} G.~J.~Ding, J.~J.~Zhu and M.~L.~Yan,
Phys.\ Rev.\ D \textbf{77}, 014033 (2008).


\bibitem{Li:2009zu} B.~Q.~Li and K.~T.~Chao,
Phys.\ Rev.\ D \textbf{79}, 094004 (2009).


\bibitem{Guo:2008zg} F.~K.~Guo, C.~Hanhart and U.~G.~Meissner,
Phys.\ Lett.\ B \textbf{665}, 26 (2008).


\bibitem{Wang:2009hi} Z.~G.~Wang and X.~H.~Zhang,
Commun.\ Theor.\ Phys.\ \textbf{54}, 323 (2010).


\bibitem{Albuquerque:2011ix} R.~M.~Albuquerque, M.~Nielsen and R.~Rodrigues
da Silva, 
Phys.\ Rev.\ D \textbf{84}, 116004 (2011).


\bibitem{Qiao:2007ce} C.~F.~Qiao,
J.\ Phys.\ G \textbf{35}, 075008 (2008).


\bibitem{Cotugno:2009ys} G.~Cotugno, R.~Faccini, A.~D.~Polosa and
C.~Sabelli, 
Phys.\ Rev.\ Lett.\ \textbf{104}, 132005 (2010).


\bibitem{Dubynskiy:2008mq} S.~Dubynskiy and M.~B.~Voloshin,
Phys.\ Lett.\ B \textbf{666}, 344 (2008).


\bibitem{Ebert:2008kb} D.~Ebert, R.~N.~Faustov and V.~O.~Galkin,
Eur.\ Phys.\ J.\ C \textbf{58}, 399 (2008).


\bibitem{Maiani:2014} L.~Maiani, F.~Piccinini, A.~D.~Polosa and V.~Riquer,
Phys.\ Rev.\ D \textbf{89}, 114010 (2014).


\bibitem{Zhang:2010mw} J.~R.~Zhang and M.~Q.~Huang,
Phys.\ Rev.\ D \textbf{83}, 036005 (2011).


\bibitem{Albuquerque:2008up} R.~M.~Albuquerque and M.~Nielsen,
Nucl.\ Phys.\ A \textbf{815}, 53 (2009) Erratum: [Nucl.\ Phys.\ A \textbf{857%
}, 48 (2011)].


\bibitem{Chen:2010ze} W.~Chen and S.~L.~Zhu,
Phys.\ Rev.\ D \textbf{83}, 034010 (2011).


\bibitem{Wang:2013exa} Z.~G.~Wang,
Eur.\ Phys.\ J.\ C \textbf{74}, 2874 (2014).


\bibitem{Wang:2016mmg} Z.~G.~Wang,
Eur.\ Phys.\ J.\ C \textbf{76}, 387 (2016).


\bibitem{Wang:2018rfw} Z.~G.~Wang,
Eur.\ Phys.\ J.\ C \textbf{78}, 518 (2018).


\bibitem{Shifman:1978bx} M.~A.~Shifman, A.~I.~Vainshtein and V.~I.~Zakharov,
Nucl.\ Phys.\ B \textbf{147}, 385 (1979).


\bibitem{Shifman:1978by} M.~A.~Shifman, A.~I.~Vainshtein and V.~I.~Zakharov,
Nucl.\ Phys.\ B \textbf{147}, 448 (1979).


\bibitem{Balitsky:1989ry} I.~I.~Balitsky, V.~M.~Braun and
A.~V.~Kolesnichenko,
Nucl.\ Phys.\ B \textbf{312}, 509 (1989). 


\bibitem{Belyaev:1994zk} V.~M.~Belyaev, V.~M.~Braun, A.~Khodjamirian and
R.~Ruckl, 
Phys.\ Rev.\ D \textbf{51}, 6177 (1995).


\bibitem{Agaev:2016dev} S.~S.~Agaev, K.~Azizi and H.~Sundu,
Phys.\ Rev.\ D\ \textbf{93}, 074002 (2016).


\bibitem{Agaev:2017tzv} S.~S.~Agaev, K.~Azizi and H.~Sundu,
Phys.\ Rev.\ D \textbf{96}, 034026 (2017).


\bibitem{Patrignani:2016xqp} C.~Patrignani \textit{et al.} [Particle Data
Group], 
Chin.\ Phys.\ C \textbf{40}, 100001 (2016).


\bibitem{Weinstein:1982gc} J.~D.~Weinstein and N.~Isgur,
Phys.\ Rev.\ Lett.\ \textbf{48}, 659 (1982).


\bibitem{Kim:2017yvd} H.~Kim, K.~S.~Kim, M.~K.~Cheoun and M.~Oka,
Phys.\ Rev.\ D \textbf{97}, 094005 (2018).


\bibitem{Agaev:2017cfz} S.~S.~Agaev, K.~Azizi and H.~Sundu,
Phys.\ Lett.\ B \textbf{781}, 279 (2018).


\bibitem{Agaev:2018sco} S.~S.~Agaev, K.~Azizi and H.~Sundu,
arXiv:1804.01726 [hep-ph].


\bibitem{Ioffe:1983ju} B.~L.~Ioffe and A.~V.~Smilga,
Nucl.\ Phys.\ B \textbf{232}, 109 (1984).


\bibitem{Agaev:2016ijz} S.~S.~Agaev, K.~Azizi and H.~Sundu,
Phys.\ Rev.\ D \textbf{93}, 114007 (2016).


\bibitem{Agaev:2016dsg} S.~S.~Agaev, K.~Azizi and H.~Sundu,
Phys.\ Rev.\ D \textbf{95}, 034008 (2017).


\bibitem{Agaev:2017foq} S.~S.~Agaev, K.~Azizi and H.~Sundu,
Phys.\ Rev.\ D\textbf{95}, 114003 (2017). 


\bibitem{Zanetti:2011ju} C.~M.~Zanetti, M.~Nielsen and R.~D.~Matheus,
Phys.\ Lett.\ B \textbf{702}, 359 (2011).
\end{thebibliography}
\end{document}